\documentstyle[aps,prl,multicol,epsf,floats,epsfig,amsmath,amssymb]{revtex}

\begin{document} 
\draft 

\title{Bose Condensation and Temperature}

\author{Enrico Celeghini$^1$ and Mario Rasetti$^2$}
%\affiliation
\address{$^1$Dipartimento di Fisica, and Sezione INFN, 
   Universit\`a di Firenze, I-50125 Firenze, Italy \\
$^2$Dipartimento di Fisica, and Unit\`a INFM,
   Politecnico di Torino, I-10129 Torino, Italy }
\maketitle 
%\date{\today}

\begin{abstract}
A quantitative analysis of the process of condensation of bosons both in harmonic traps 
and in gases is made resorting to two ingredients only: Bose classical distribution and 
spectral discretness. It is shown that in order to take properly into account statistical 
correlations, temperature must be defined from first principles, based on Shannon entropy, 
and turns out to be equal to $\beta^{-1}$ only for $T > T_c$ where the usual results are 
recovered. Below $T_c$ a new critical temperature $T_d$ is found, where the specific heat 
exhibits a sharp spike, similar to the $\lambda$-peak of superfluidity. 
\end{abstract}
\date{today}

\begin{multicols}{2}
\narrowtext 

In the analysis of low temperature behaviour for Bose systems, temperature is typically 
thought of as given by $T= 1/ k_B \beta$, $\beta$ being the Lagrange multiplier one has to 
introduce in the customary ensemble approach to account for energy conservation. Such 
definition however implies in fact the assumption of statistical independence of modes, 
which not only is not necessary but may even be in contradiction with the basic features 
of the conceptual scheme leading to the construction of quantum statistics. Indeed, the 
latter rests on the notion of density matrix, which is correctly introduced only after 
the Fock space has been properly generated over the tensorized product of the appropriate 
number of single-particle states. This, in turn, implies the existence of mode correlations 
which are there even in the absence of interactions. Only when temperature is high enough 
as to break such correlations, the Fock density matrix can be lifted to a product of single 
particle density matrices, in which case, however, the only statistics achievable is 
Maxwell-Boltzmann's, where correlations are absent. 

Ever since 1907 Einstein himself \cite{Ein907}, discussing the behavior at low temperature 
of the quantized harmonic oscillator in the frame of Boltzmann statistics, pointed out how 
classical results on equipartition of energy are, in agreement with the Nernst theorem, no 
longer valid for $T < h \nu /k_B$, by obtaining a good fit of the heat capacity data for 
diamond. One of his basic assumptions was the extensivity hypothesis. The latter 
is usually transferred, adopting a tensor product structure for the Hilbert space of states 
of a multi-particle system, to the whole Quantum Statistical Mechanics (QSM) in spite 
of the fact that it can be justified for the Gibbs-Boltzmann distribution only, related 
in this case to the statistical independence of different modes. It worked well for 
Einstein when he dealt with the carbon atoms in diamond (which of course do not condense) 
in that, being fixed at their positions in the crystal, these are distinguishable, while 
one of the characteristic features of quantum mechanics is indistinguishability of the 
particles. 

Several authors have discussed Bose Einstein condensation (BEC) in magnetic traps, all on 
the basis of that same definition of temperature \cite{Ensher},\cite{BPK},\cite{dHT}, 
finding -- together with $T_c \propto N^{1/3}$ -- a heat capacity monotonically decreasing 
to zero for $T$ below $T_c$, in agreement with what happens for true gases under the 
same hypotheses. 

On the other hand Bose, when he $''$invented$\, ''$ bosons, broke scale invariance in the 
same way as Democritus did in the 5th century BC: because of correlations, a condensate is 
not the disjoint union of smaller condensates, in the same way as an atom (ancient or 
modern) is not equal to two half-atoms. The properties of a condensate do depend on the 
number $N$ of atoms it contains, as one cannot consider the thermodynamic limit \cite{KD}
\cite{GH}, \cite{KT}, \cite{DGPS} that would require, together with $N \to \infty$, $h 
\nu \propto N^{-1/3}$ ({\sl i.e.} the {\sl continuum}), in conflict with the properties of 
magnetic traps where the spectrum ({\sl discrete}) has a spacing which is fixed and 
comparable with $k_B T_c$. In other words the distinction of physical quantities in 
intensive and extensive is not consistent with Bose-Einstein condensation, as proved, 
among others, by the mentioned relation $T_c \propto N^{1/3}$.    

All these well known features are little relevant for $T > T_c$, because the particles 
there are distributed over all levels, and the continuum approximation appears to be 
sufficient both for describing the regime for $T > T_c$ and for determining $T_c$ itself 
(minor corrections may arise, that will be considered elsewhere \cite{CeRa}). However, as 
soon as the condensate fraction $c \equiv \overline{n}_0/N$ becomes relevant ($T < T_c$), 
statistics becomes significantly different from Boltzmann's and the very definition of 
temperature must be reconsidered together with, but independently from, the question of 
equipartition. Temperature can be defined, in a quite general and basic way, in terms of 
two fundamental physical quantities; internal energy $E$ and entropy $S$. Thus instead of 
connecting $T$, because of extensivity, to the Lagrange multiplier $\beta$, one can 
derive it from the separate measurements of $E$ and $S$.

We shall then consider {\sl ab initio}, in the grand-canonical ensemble, a system of $N$ 
bosons of total energy $E$ and we shall explicitly evaluate the Lagrange multipliers as 
functions of $N$ and $E$. Successively, Shannon's form of entropy will be our starting 
point to derive all equilibrium thermodynamic properties. In other words, our hypotheses 
will be the three consistent assumptions: \\
a) The Bose-Einstein statistics:
\begin{eqnarray}
\overline{n}_i = \frac{1}{e^{\alpha + \beta \epsilon_i}-1} \; , \label{BE}
\end{eqnarray}
b) a discrete energy spectrum (as suggested by the typical experimental set-up with 
magnetic traps, where $k_B T_c \approx 10\, h\nu$), and\\ 
c) Shannon entropy, evaluated in terms of $N$ and $E$. \\ 
Temperature $T = T(N,E)$ will then be obtained as a derived quantity by $T=\partial 
E/\partial S$.     

%%%%%%%%%%%%%%%%%%%%%%%%%%%%%%%%%%%%%%%%%%%%%%%%%%
%\begin{figure1}
\hspace{1cm}

% \begin{figure}
\begin{center}
\epsfig{file=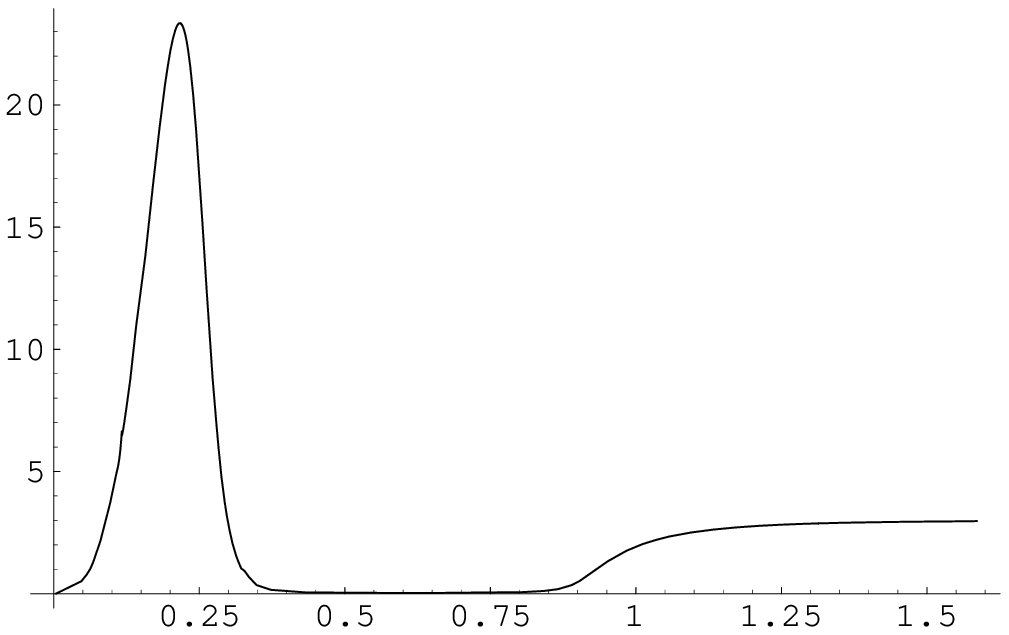,height=5cm} 
\end{center}

% \caption 
{\noindent{\footnotesize{Fig. 1. Specific heat $C/k_B$ 
{\sl vs.} $T/T_c$ for $N=10^6$ bosons in an isotropic harmonic trap, as predicted by 
the entropy approach.}}}

% \end{figure}
\hspace{1cm}

%end{figure1}
%%%%%%%%%%%%%%%%%%%%%%%%%%%%%%%%%%%%%%%%%%%%%%%%%

The resulting thermodynamics of bosons at temperatures above and below $T_c$ is thus 
simply derived numerically, both for an harmonic potential and a cubic box, from the two 
inputs $N$ and $E$. The temperature scale $T_c$ is found to confirm the results of 
reference \cite{DGPS}, the specific heat $C/k_B$ has the appropriate value for $T > T_c$; 
3 for the three dimensional harmonic oscillator and 3/2 for the gas in a box (as it should, 
since the system does not condense and behaves essentially as a classical gas); however 
new features appear for $0 < T < T_c$. For both the physical situations considered, heat 
capacity exhibits an unexpected behaviour; at $T_c$, it falls briskly down to almost zero 
(entropy and energy decrease with quite different slopes because only a few atoms migrate  
progressively from the warmer tail of the spectrum into the ground level) and raises up 
abruptly at a new critical temperature $T_d$ (where $\overline{n}_0/N \approx 1/2$), at 
which it exhibits a sharp spike, analogous to that of superfluids \cite{Ah}. As no 
interaction has been introduced, other than the confining potential, this enhance the idea 
that superfluid transition may indeed be ascribable, as originally suggested by Onsager and 
Penrose \cite{OP}, essentially to Bose-Einstein condensation. For $T < T_d$, specific heat 
goes to zero when $T$ goes to zero, in agreement with Nernst's theorem. It is suggestive 
that, for $T\to 0$, $C \propto T^3$ for harmonic traps whereas $C \propto T^{3/2}$ for 
bosons in a box. In Fig.1 $C/k_B$ is given {\sl vs.} the $''$probabilistic$\, ''$ 
temperature. To check the procedure, $C$ has also been numerically evaluated as 
$\partial E/\partial \beta^{-1}$, checking that this coincides with the standard textbook 
form (see, {\it e.g.}, \cite{Bal}). It is thus clear that the effect lies in the different 
definitions of temperature and not in computations. 

We shall now summarize a few technicalities of the procedure, emphasizing that 
the scheme proposed to deal with correlations could in principle be utilized also for 
interactions.   

1) A set of values for $N$ in the range $10^6 \div 10^9$ as well as a set of values for 
the mean energy per particle $E/N$ in the range from the ground state up to $10^4$ units 
$h\nu$ for the magnetically trapped system and $h^2/L^2$ for the cubic box (of side $L$) 
have been considered. The difference between the two cases lies in the energy spectrum and 
its level multiplicity. The full range of condensation $0 < c \equiv \overline{n}_0/N < 1$ 
has thus been covered.  

2) For each pair of values $N$ and $E$ the system of two equations in the two unknowns 
$\alpha$ and $\beta$:    
\begin{eqnarray}
\sum \,\overline{n}_i(\alpha ,\beta) = N \quad , \quad 
\sum \, \epsilon_i \,\overline{n}_i(\alpha ,\beta ) = E \; , 
\nonumber 
\end{eqnarray}
(in the sums the correct state multiplicities have of course been taken into account) was 
solved. The two functions $\alpha = \alpha(N,E)$ and $\beta = \beta(N,E)$ were thus found.

3) Successively, the $\overline{n}_i = \overline{n}_i(N,E)$ were obtained from eq. 
(\ref{BE}). In particular the condensate fraction $c \equiv \overline{n}_0/N$ was 
determined as a function of $N$ and $E$.

4) The probability of finding a particle in state $i$, picking it up at random in the 
ensemble is $p_i(N,E) = \overline{n}_i/N$. This allows us, adopting Shannon's definition, 
to find the entropy $S$ as a function of $N$ and $E$; 
\begin{eqnarray}
S = - k_B \, \sum \, p_i \, \ln p_i . \nonumber 
\end{eqnarray}
In this way the correlations induced by quantum statistics \cite{Bal} enter into play 
only in the definition of the $p_i$'s which are such that $p_i^{(1,2)} \neq p_i^{(1)} 
p_i^{(2)}$ if the system is split into the union of two parts. It is interesting 
to note that, had we considered Boltzmann distribution and not eq.(\ref{BE}), Gibbs'  
distribution would have been obtained and the results of \cite{Ein907} reproduced. 
Once the $p_i$'s are known, any further analysis 
of equilibrium thermodynamical properties is mere statistics. 

5) Temperature $T$ is obtained, as a function of $N$ and $E$, by numerical derivation: 
\begin{eqnarray}
T \equiv T(N,E) = \left [\frac{\partial S}{\partial E}\right ]_N^{-1} \; . 
\nonumber 
\end{eqnarray} 

6) The relation $T = T(N,E)$ is monotonic in $E$ for fixed $N$, and it can be inverted to 
obtain $E = E(N,T)$, whence the specific heat is obtained by (numerical) differentiation (see Fig. 1): 
\begin{eqnarray}
C \equiv  \left [\frac{\partial (E/N)}{\partial T}\right ]_N \; . 
\nonumber 
\end{eqnarray} 
\indent 7) $''$Probabilistic$\, ''$ temperature $T$, despite its conceptual relevance, is 
perhaps not the physical parameter that best describes the process of condensation: it is 
difficult to measure in the experimental set-up of traps and, as one does not have 
equipartition, its intuitive meaning of $''$average energy per particle$\, ''$ is lost. 
On the other hand, experimentalists define temperature deriving it from the high energy 
tail of the velocity distribution for which they assume \cite{Ensher} equipartition. In 
other words, the $''$experimental$\, ''$ temperature is 
\begin{eqnarray}
  {\Theta} := \frac{1}{3 k_B} \frac{1}{N-\overline{n}_0} \sum_{j\neq 0} \overline{n}_j \epsilon_j \; . 
  \nonumber 
\end{eqnarray}   
We therefore describe the onset of condensation by exhibiting the behavior of $c$ {\sl 
vs.} ${\Theta}$ (Fig. 2), that we compare with $c$ {\sl vs.} $T$ (Fig. 3). 

%%%%%%%%%%%%%%%%%%%%%%%%%%%%%%%%%%%%%%%%%%%%%%%%%%
%\begin{figure2}
\hspace{1cm}

\begin{center} 
% \begin{figure}
\epsfig{file=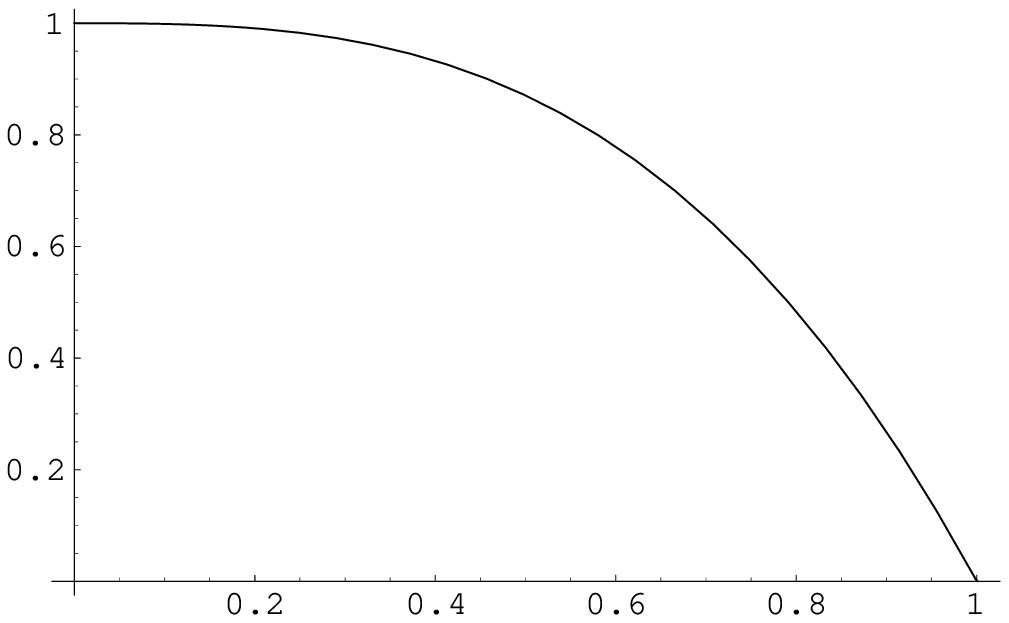,height=5cm}
\end{center} 

% \caption 
{\noindent{\footnotesize{Fig. 2. Condensate fraction $c\equiv\overline{n}_0/N$ {\sl 
vs.} $''$experimental$\, ''$ temperature ${\Theta}/T_c$. The curve is perfectly 
fitted by $c=1-\bigl (\Theta/T_c\bigr )^3$.}}} 

% \end{figure}
\hspace{1cm}

%end{figure2}
%%%%%%%%%%%%%%%%%%%%%%%%%%%%%%%%%%%%%%%%%%%%%%%%%

For the sake of completeness, we report also the $''$experimental$\, ''$ specific heat 
${\cal C} \equiv \partial \bigl ( E/N \bigr ) /\partial\Theta$ (Fig. 4). It should 
be noticed that since above critical temperature equipartition holds also in our case, 
the two critical temperatures coincide (${\Theta}_c \equiv T_c \approx .9 h\nu/k_B 
N^{\frac{1}{3}}$) and equal that given in \cite{Ensher}. In Fig. 5 the specific heat of 
the $''$warm tail$\, ''$, ${\cal C}_w \equiv \partial (3 k_B \Theta )/\partial T$ is shown 
{\sl vs.} $T/T_c$).    

8) As we have the complete description of the system in function of $N$ and $E$, all 
physical observables can be obtained. For instance the chemical potential, $\mu \equiv 
\mu(N,E) = \bigl ( \partial E / \partial N \bigr )_S$, which gives $\alpha \neq -\beta\mu$. 

In order to make the figures more easily comparable with known results in the literature, 
only those referring to a single value of $N$ ($N = 10^6$) and to the harmonic confining 
trap are reported. It should be pointed out that, taking into account that $T_c \propto N^{1/3}$,
no substantial differences appear either for different values of $N$ nor for the gas 
in a box. 

%%%%%%%%%%%%%%%%%%%%%%%%%%%%%%%%%%%%%%%%%%%%%%%%%%
%\begin{figure3}

\hspace{1cm}

% \begin{figure}
\begin{center} 
\epsfig{file=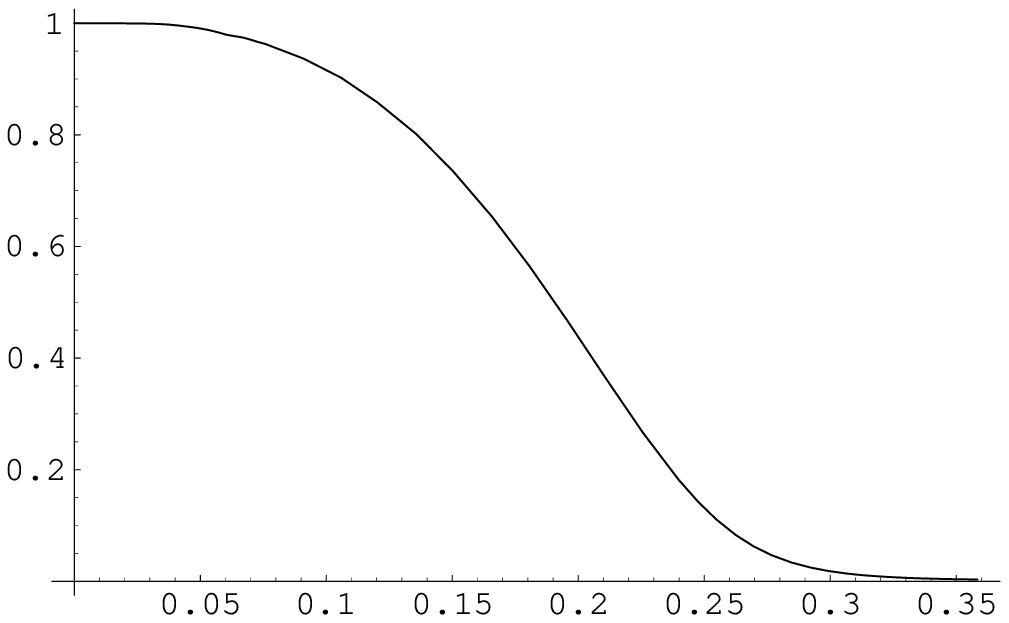,height=5cm}
\end{center} 

% \caption 
{\noindent{\footnotesize{Fig. 3. $c$ {\sl vs.} $T/T_c$ ($T$ probabilistic temperature). 
Apparently similar to Fig. 2, just merging to zero in a smoother way, this is indeed 
deeply different, in that the critical temperature is $T_d$ instead of $T_c$}}}

% \end{figure}
\hspace{1cm}

%end{figure3}
%%%%%%%%%%%%%%%%%%%%%%%%%%%%%%%%%%%%%%%%%%%%%%%%%

%%%%%%%%%%%%%%%%%%%%%%%%%%%%%%%%%%%%%%%%%%%%%%%%%%
%\begin{figure4}
\hspace{1cm}

\begin{center} 
% \begin{figure}
\epsfig{file=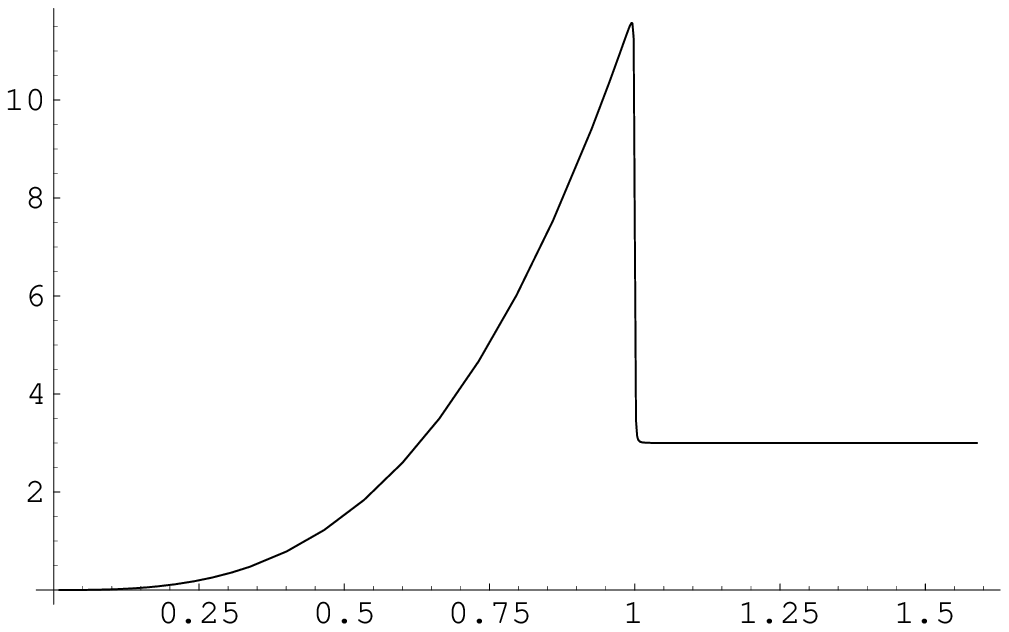,height=5cm}
\end{center} 

% \caption 
{\noindent{\footnotesize{Fig. 4. $''$Experimental$\, ''$ specific heat ${\cal C}/k_B\equiv\partial 
(E/N)/\partial(k_B \Theta)$ {\sl vs.} $\Theta/T_c$. Well reproduces the result in \cite{KD}.}}} 

% \end{figure}
\hspace{1cm}

%end{figure4}
%%%%%%%%%%%%%%%%%%%%%%%%%%%%%%%%%%%%%%%%%%%%%%%%%

Our interest is focussed on different aspects. On the one side the proposed combinatorial 
approach emphasizes the relevance, for a proper interpretation of phenomenological data, 
of a definition of temperature adequate to the experimental set up. On the other, it hints 
to the existence at very low temperature of effects that would be worth investigating in 
the laboratory.  Such effects are probably at present beyond reach, but experiments are 
in constant evolution, and the effort might open interesting new perspectives. Finally, it 
may provide a novel insight on the $\lambda$-transition in superfluids. 

QSM has always been at the border between statistical mechanics (SM) and probability 
theory. For the latter \cite{JKK} SM is related to information theory, where also the lexicon is 
different from that usual one in physics. For instance, $''$Boltzmann statistics$\, ''$ is
simply the common limit for $\overline{n}_i \ll 1$  of the only two existing statistics, Bose
and Fermi in \cite{aaa}, while is just one possible $''$physical distribution$\, ''$ with
unlimited occupation numbers, like that of Bose in \cite{JKK}, \cite{CR}. 

%%%%%%%%%%%%%%%%%%%%%%%%%%%%%%%%%%%%%%%%%%%%%%%%%%
%\begin{figure5}
\hspace{1cm}

\begin{center} 
% \begin{figure}
\epsfig{file=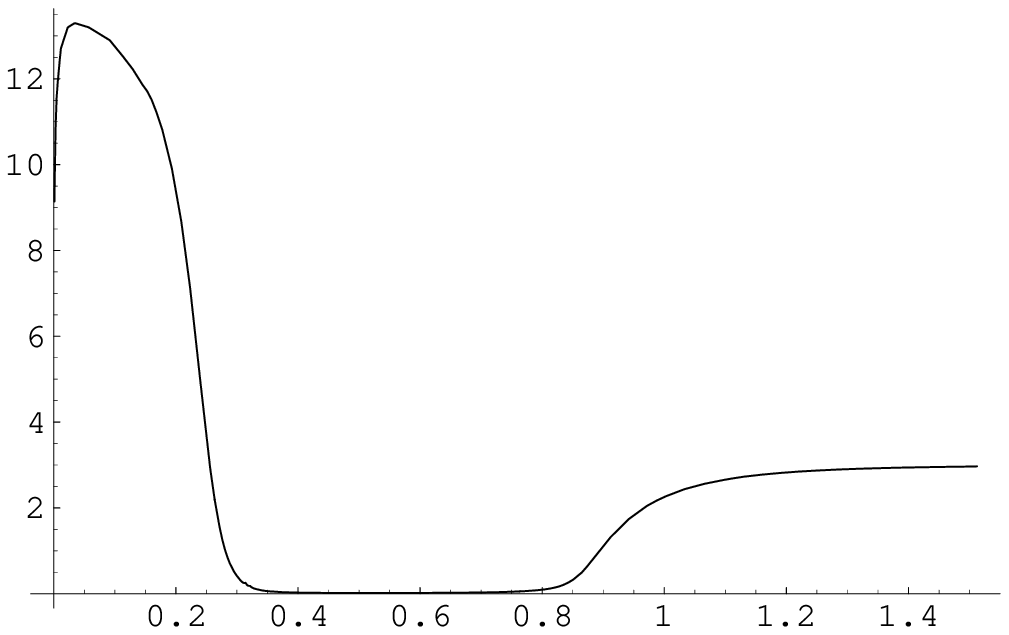,height=5cm}
\end{center} 

% \caption 
{\noindent{\footnotesize{Fig. 5. Specific heat of the warm tail ${\cal C}_w/k_B$ {\sl vs.} 
$T/T_c$. ${\cal C}_w/k_B$ is constant ($= 3$) for 
$T>T_c$ as expected classically, and it is two orders of 
magnitude smaller for $T_d<T<T_c$, and $\sim 4$ times larger for $0<T<T_d$.}}} 

% \end{figure}
\hspace{1cm}

%end{figure5}
%%%%%%%%%%%%%%%%%%%%%%%%%%%%%%%%%%%%%%%%%%%%%%%%%

In distribution theory all information is contained in weights such as those in eq. 
(\ref{BE}), while the fundamental hypothesis of statistical mechanics is statistical 
independence of the subsystems. In QSM it is the symmetrization imposed by particle 
indistinguishability that one has to perform, {\sl e.g.} moving to Fock space, that breaks 
such independence.  Results of physical interest are nevertheless typically in agreement, 
which is due in most cases to the fact that statistical correlations are irrelevant and 
therefore all physical distributions are equivalent. Well established exceptions are fermions 
at low temperature and, referring to bosons, the black body, where photons are well 
described by eq.(\ref{BE}). In black body theory, however, the situation is very peculiar 
because photons essentially cannot interact with each other. 

On the other hand, in view of the continuous improvements of experimental techniques,  
extending QSM to describe experiments where $\overline{n}_i \not> 1$, the spectrum is 
discrete, the number of particles and modes is large but not so large as to allow a 
continuum description, is certainly a challenge that will have to be faced soon. Indeed 
all of this is already happening just for one physical set-up: Bose condensation in 
magnetic traps. 

In the condensate most particles are in the fundamental level and, in typical experimental 
situations, $k_B T_c$ is only of the order of ten times the spectrum spacing ($T_c \approx 
\, 10^2 nK$, $h\nu /k_B \approx 10\, nK$). The usual approximation in terms of continuum 
density with $\overline{n}_0$ obtained by difference, appears to be improvable. Moreover, 
the statistical independence of levels -- acceptable in the black body -- is questionable 
when the container is a magnetic field and has no kind of thermal exchange with the atoms. 
Thermodynamic equilibrium of each mode is reached because not of the interaction of the 
mode itself with the bath, but of the jumping of atoms from one level to another, as 
clearly shown by the fast restoring of equilibrium after each cooling. In the case of gas, 
temperature is controlled by the walls and the physics of thermalization is different but, 
if the statistics is that of Bose, statistical independence must be excluded.

Keeping the spectral structure into account provides the correct conceptual scheme as  
opposed to a theory where continuum and equipartition are assumed {\sl a priori}. On the 
other hand, many body effects so far not considered should be analysed; however we expect 
that including in the picture interactions in the ground state, as described in the 
Gross-Pitaevsky approach \cite{GSS}, \cite{DGPS}, may modify at most the energy spectrum, 
and the ensuing differences in numerical values should not be able to substantially mask 
the effects discussed. 

It should be pointed out that the description presented, based on the assumption of 
spectral discreteness, leads not only to qualitative behaviours, but it determines as well 
physical observables: specific heat, filling order parameter, chemical potential, etc., all 
of which can be experimentally measured, related to the characteristics of the experiment 
($N$, $E$, $\nu$). To this effect, it should be observed that in real experiments the 
harmonic potential is usually anisotropic: extension of our results to this situation is 
straightforward and leads to quite similar predictions. 

% In conclusion, this paper aims to be both a contribution to a deeper understanding of 
% the profound relations between combinatorics and extensive hypothesis on one side, and 
% an exhaustive description of condensed bosons below $T_c$ -- that we propose to 
% experimentalists for a check -- on the other.

\end{multicols}
\end{document}